\documentclass[12pt]{article} 
\usepackage[dvips]{graphicx}

\begin{document}


\centerline{\Large\bf{Study of the Nuclear Activity of the Seyfert Galaxy }}
\centerline{\Large\bf{NGC 7469 over the Period of Observations 2008--2014}}

\bigskip

\centerline{L.~Ugol'kova$^{1*}$, B.~Artamonov$^1$, E.~Shimanovskaya$^1$,}
\centerline{V.~Bruevich$^1$, O.~ Burhonov$^2$, Sh.~Egamberdiev$^2$, N.~Metlova$^1$}
\bigskip
\centerline{{\it $^1$ M. V. Lomonosov Moscow State University, Sternberg Astronomical Institute, Moscow}} 
\centerline{{\it $^2$ Ulugh Beg Astronomical Institute of the Uzbek Academy of Sciences, Tashkent}}

\vspace{2mm}

\sloppypar 
\vspace{2mm}
\noindent

{\bf Abstract.}
We present results of multicolor UBVRI observations of the type 1 Seyfert
galaxy (SyG 1) NGC 7469 carried out at the 1.5-meter telescope of the Maidanak Observatory (Uzbekistan) in 2008--2014. Analysis of the light curves indicates the presence of another slow flare of a long-term variability in 2009-2014 with a maximum in 2011-2012. We investigate properties of the long-term variability in 2009-2014, present (U-B)--(B-V) color diagrams for maxima and minima of  NGC 7469 nuclear variability using various apertures and compare them with the black-body gas radiation which models the accretion disk radiation. Color-index measurements shows that the color  becomes bluer at maximum brightness, indicating a higher
temperature of the accretion disk.
We have analysed the relation of X-ray and optical variability of NGC 7469 in 2008 and 2009 in
comparison with the activity minimum in 2003. In 2008 the correlation coefficient between the X-ray and optical radiation is close to 0.5. Such poor correlation can be explained by the influence of an SN 1a explosion in close proximity to the nucleus of NGC 7469. The SN manifests itself in the optical band but does not affect the X-ray variability pattern.
Comparison of the variability data in 2009 reveals a good correlation between the optical (U band) and the X-ray (7-10 keV) variability with the correlation coefficient of about 0.93. The correlation coefficient and the lag depend on the wavelength in the optical and X-ray bands. The lag between the X-ray and optical fluxes in 2009 is 2--4 days. In 2003 the lag is almost zero.


\clearpage

\section*{Introduction}
Extensive observations of active galactic nuclei showed that AGNs vary in all wavelength bands on different time scales. Study of the variability can provide information about structure and energy-generation mechanisms in AGNs.
Long-term observations of the Seyfert galaxy NGC~7469 (Arp 298 = MCG 1-58-25) revealed the presence of variability with time scales from minutes to decades. 

NGC~7469 is a spiral galaxy of the SBa type, slightly inclined to the line of sight. The object coordinates are RA 23h 03m 15.75s, DEC $+08œ$ $52'$ $25''.9$. The distance to the galaxy is $D=68$ Mpc for $H_0=75$ km/s/Mpc, $z = 0.01639$. A physical companion, the irregular galaxy IC 5283, is situated at the distance of $80''$. In IR and optical bands, a star-burst region is observed around the nucleus in the form of a ring with the diameter of $1.5-2.5''$.

The central part of the galaxy is variable in the X-ray, UV, optical and IR ranges. In the radio-frequency range, NGC~7469 is a weak source. The variability is also observed in spectral lines. In the optical range, the photometry of NGC~7469 was performed by Doroshenko, Lyuty and Rahimov (1989) in 1967--1987. The variability is confirmed in observations by Merkulova  (2000) in 1990--1998 and Sergeev, Doroshenko et al. (2005, 2010) in 2002--2009 at the Crimean Astrophysical Observatory. Results of Maidanak observations of NGC~7469 in 1990--2007 are presented by Artamonov et al. (2010).

As for many other AGNs, the variability of the NGC~7469 nucleus can be represented as a superposition of a slow component with a duration of several years and a flare, or fast, one with a duration from several days to tens of days. We will refer them as S and F components, respectively. That model was suggested by  
Lyuty and Pronik (1975, 2006). The maximum of the S component is usually attributed to an active state of a Seyfert galaxy nucleus. The F component is probably connected to an accretion rate.

The lightcurve of NGC~7469 over observational period 1990--2014 has two stages of the long-term variability: in 1994--2001 with a maximum radiation in 1997-1998, and in 2003--2007 with a maximum radiation in 2005. Combined lightcurves of NGC~7469 based on various observations in 1990--2007 were published in the proceedings of the Odessa Gamov Conference in 2011 (Ugol'kova, Artamonov (2011)). Chesnok et al. (2009) investigated the optical-X ray cross-correlation in the activity minimum in 2003 and discovered almost zero time delay between X-ray emission with respect to optical variability in the B band. Doroshenko et al. (2010) presented lightcurves of NGC 7469 based on Crimean observations and calculated time delays between optical and X-ray emission in different activity periods, they found the correlation with optical variability in B band in 2003--2007.

\section*{Observations}
\noindent

In 2008--2014, CCD observations of NGC~7469 were carried out at 1.5-m telescope AZT--22 of the Maidanak Observatory (Ulugh Beg Astronomical Institute of the Uzbek Academy of Sciences) during 1.5-3 months every year. The  SNUCAM 4096È4096 CCD array and UBVRI filters of the Cousin's system were used. The frames were processed with the reduction software developed based on ESO-MIDAS (European Southern Observatory Munich Image Data Analysis System) package. 
The main image reduction stages were correction for bias and flat field, removal of cosmic-ray traces, determining the sky background, then subtracting it from each image frame. 
Photometry measurements were taken with apertures $10''$, $15''$, $20''$ and $30''$. As far as the intra-night variability goes beyond the scope of this work, the data were averaged for every night. Reference stars are the same as in the article by Doroshenko et al. (2005), they are on the same frame with the object, so the atmospheric extinction and variable air mass effects were not taken into account. Typical uncertainties of the CCD observations are in the range from 0.001 to 0.01 mag. The lightcurves of NGC 7469 nucleus variability in 2008--2014 are presented in Fig. 1,2.  
We did not subtract the constant galaxy background because it is model dependent.

Analysis of the long-term variability of NGC~7469 in 1990--2007 (Artamonov et al. (2010)) and in 2008--2014 revealed the presence of one more cycle of variability of the S component with duration of 6 years (2009--2014) and fast flux changes of the F component with durations from several days to 2 months. An amplitude of the fast variability (F component) is always less than that of the S component. The smaller a flare duration is, the smaller a flare amplitude is. The same conclusion is true for the S-component too. The brightness of NGC~7469 in 2009--2014 activity cycle is a bit weaker, than in 1994--2001, but is higher than in the 2004--2007 activity cycle. A maximum of that S component cycle is between 2010 and 2012.

\begin{figure}[tbh!]

\center{\includegraphics[scale=1.2]{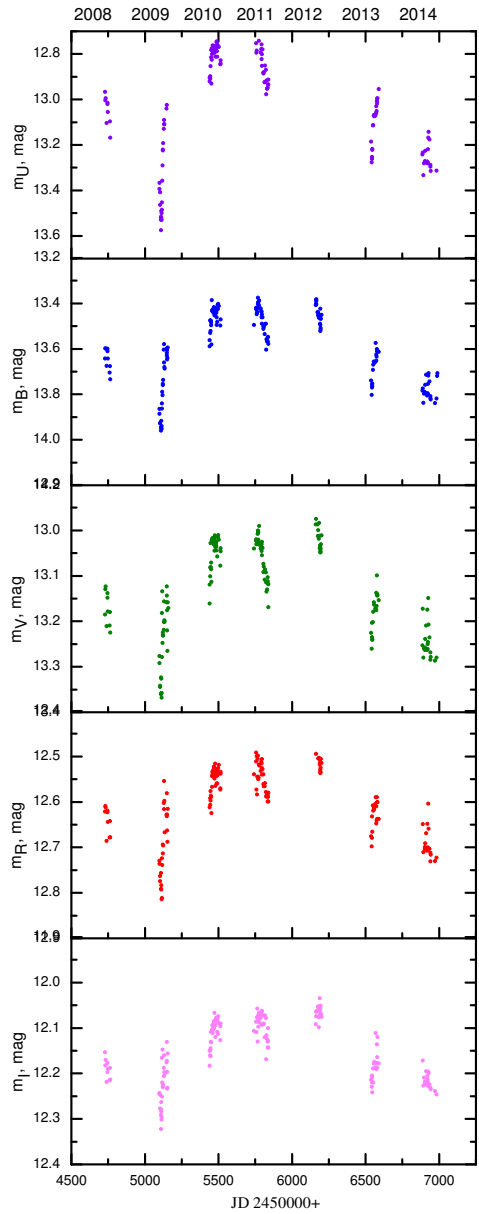}} 
\caption{Light curves of NGC~7469 in 2008--2014: U, B, V, R, I passbands; the aperture $d=10''$.}
{\label{UBVRI20082014}}
\end{figure}

\begin{figure}[tbh!]
\begin{minipage}[h]{0.23\linewidth}
\center{\includegraphics[width=1\linewidth]{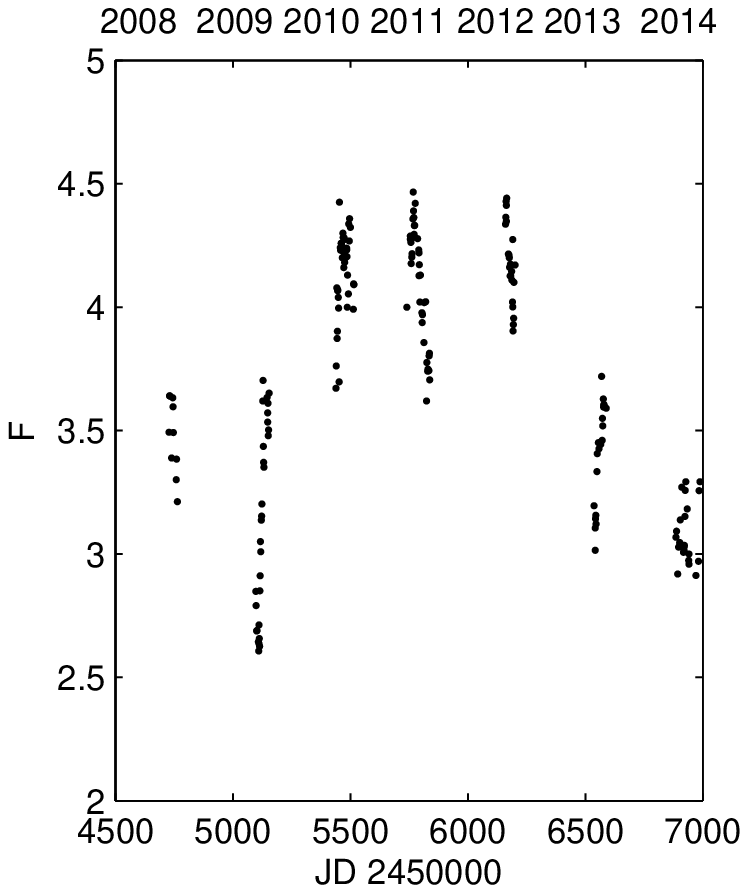}} \\
\end{minipage}
\begin{minipage}[h]{0.23\linewidth}
\center{\includegraphics[width=1\linewidth]{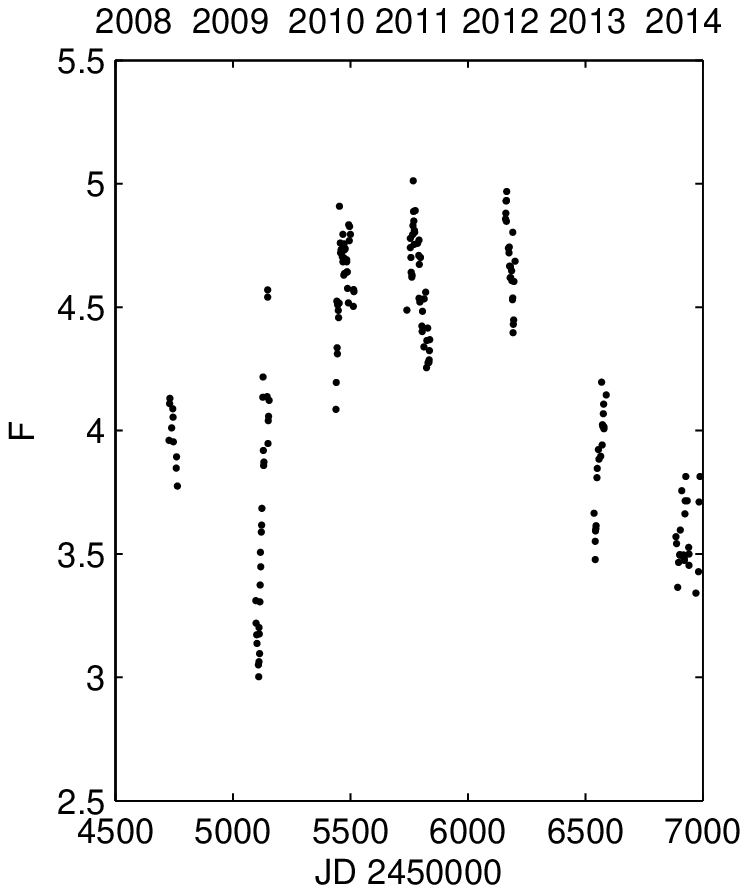}} \\
\end{minipage}
\begin{minipage}[h]{0.23\linewidth}
\center{\includegraphics[width=1\linewidth]{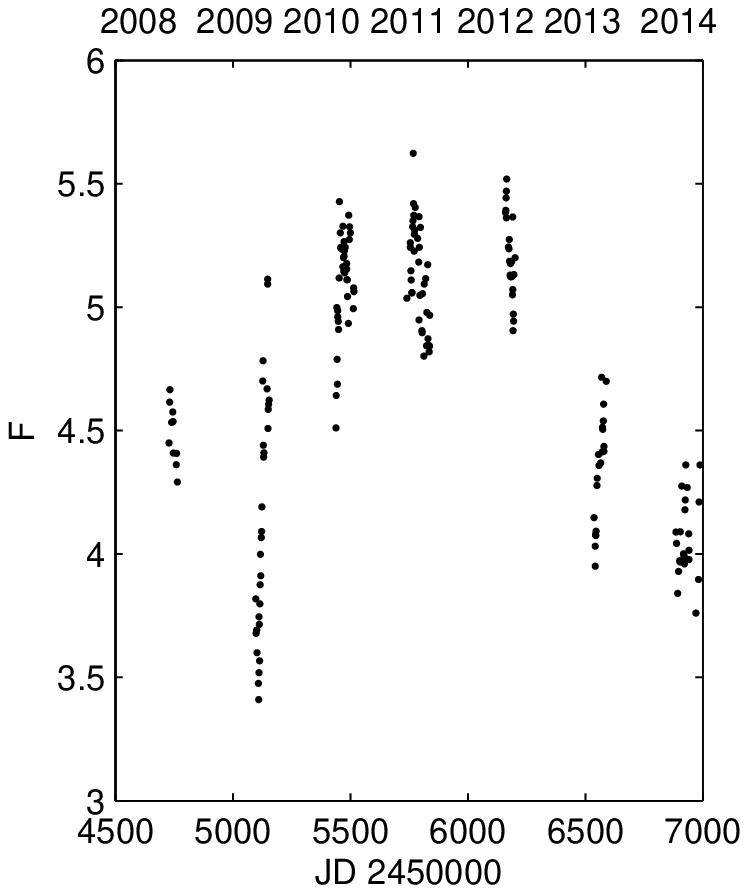}} \\
\end{minipage}
\begin{minipage}[h]{0.23\linewidth}
\center{\includegraphics[width=1\linewidth]{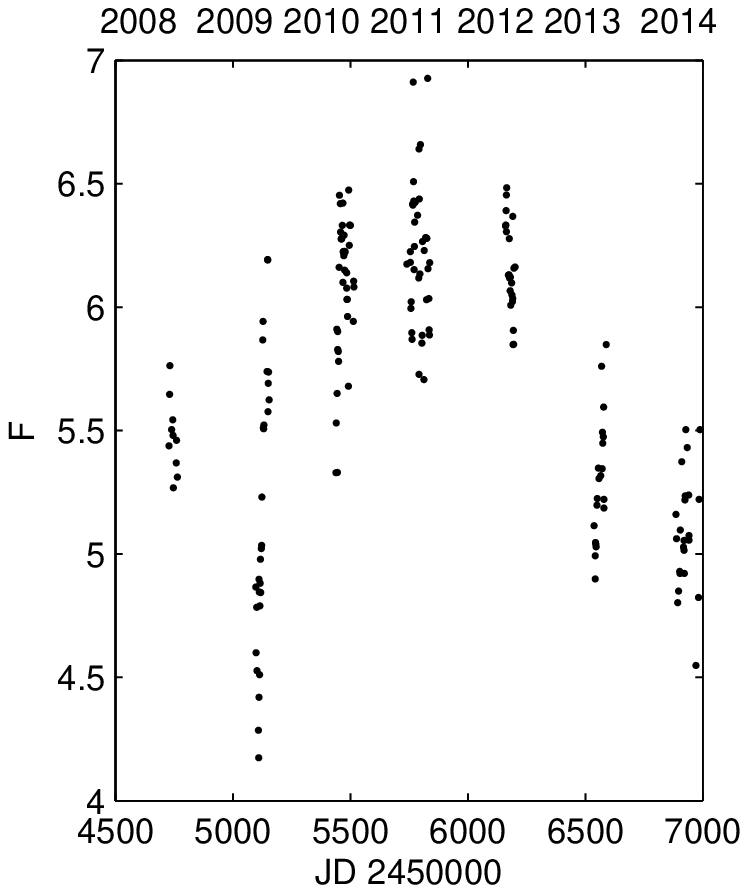}} \\
\end{minipage}
\vfill
\begin{minipage}[h]{0.23\linewidth}
\center{\includegraphics[width=1\linewidth]{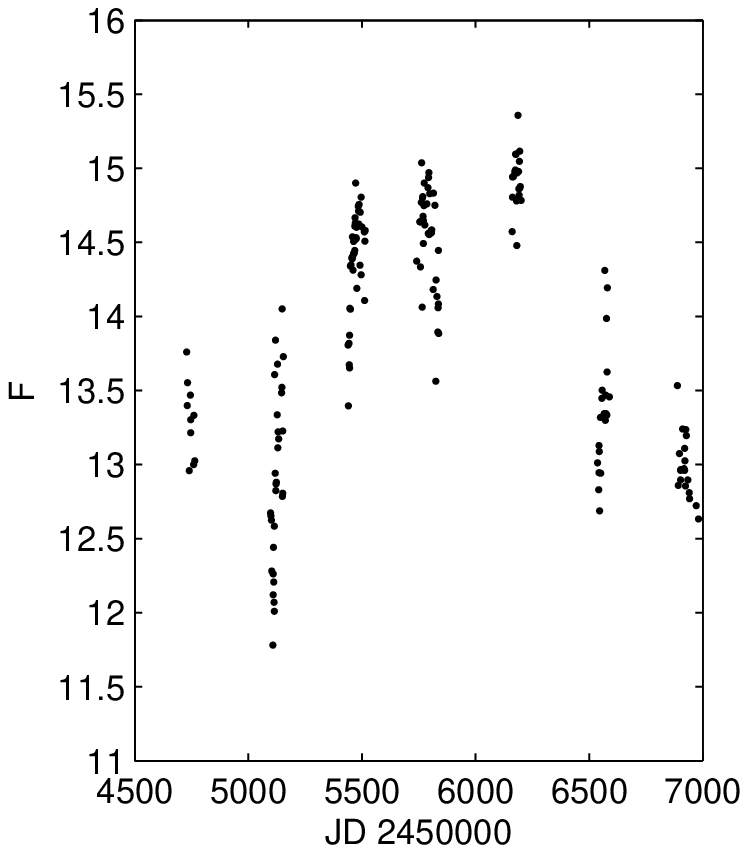}} \\
\end{minipage}
\begin{minipage}[h]{0.23\linewidth}
\center{\includegraphics[width=1\linewidth]{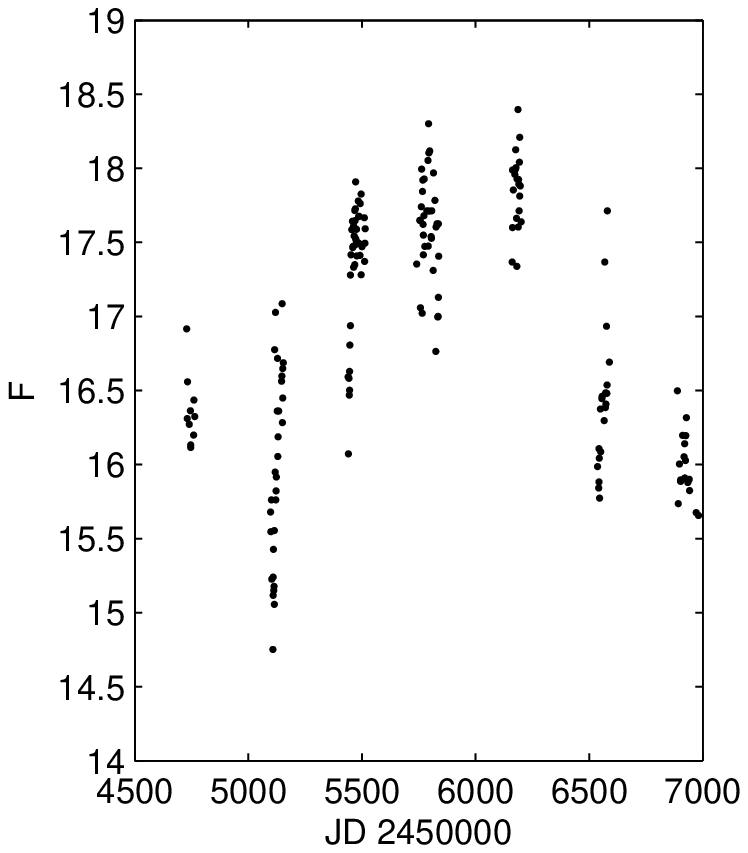}} \\
\end{minipage}
\begin{minipage}[h]{0.23\linewidth}
\center{\includegraphics[width=1\linewidth]{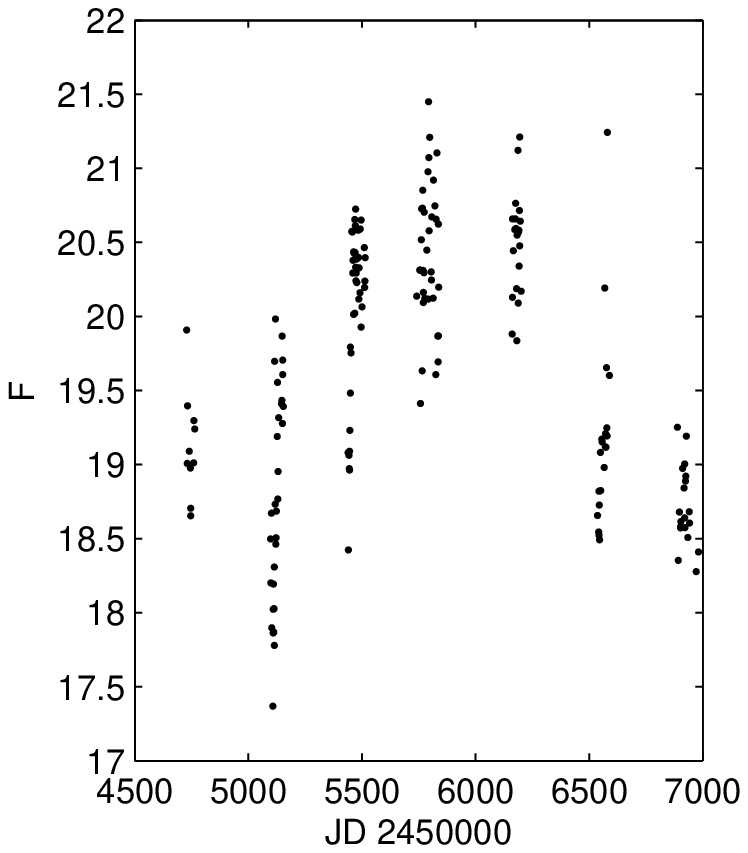}} \\
\end{minipage}
\begin{minipage}[h]{0.23\linewidth}
\center{\includegraphics[width=1\linewidth]{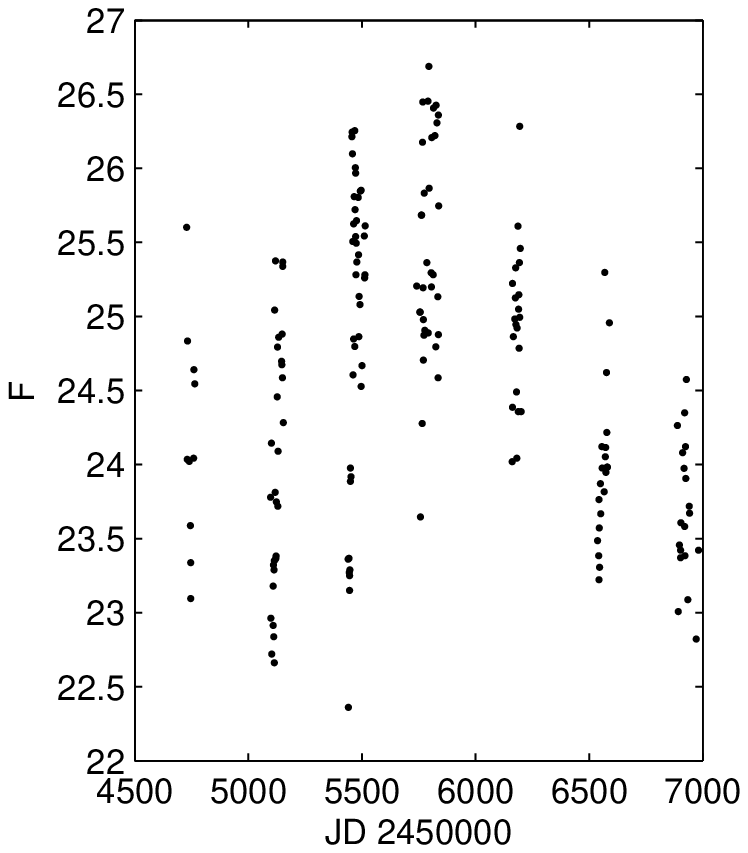}} \\
\end{minipage}
\caption{The light curves in B (top) and I (bottom) passbands (in Fk flux units), where $k=10^{15}$ $erg cm^{-2} s^{-1} A^{-1}$ from 2008 till 2014 for four apertures: $d=10$, $15$, $20$, and $30$ arcsec).}
\label{BIlc}
\end{figure}

The UBVRI photometry (the aperture $d=10''$) of NGC~7469 is presented in Fig. 1 for the monitoring period 2008--2014. The duration of the rise part is shorter than that of the descending part of the light curves in the U and B bands. The lightcurve gradient rise from the I band to the U band. The variability amplitude also increases from the I band to the U band, that is suggestive of indicative emission burst in the blue part of the spectrum in active galactic nuclei. This is most clearly seen in Fig. \ref{BIlc}, where B and I light curves in relative flux units in 2008--2014 are presented for apertures with different diameters (10, 15, 20 É 30 arcsec). For apertures with $d>15''$, the brightness maximum occurs in 2011 in all bands, ascending and descending arms are symmetrical in all bands (see Fig.\ref{BIlc}).
An area, where brightness variations of a central part of the galaxy occur, is much less than the smallest aperture diameter ($10''$) that we use for obtaining photometry. We did not investigate variability in apertures with $d<10''$, as the PSF analysis based on stars in the field of the galaxy suggested that a size of PSF wings is about $5-7''$. The contribution of wings is negligible, but it can still lead to photometry errors when measuring the star-like nucleus of the galaxy.

For more detailed analysis of NGC~7469 variability, we present 
the galaxy UBV brightness in apparent magnitudes for 
the aperture $d=10''$ in time windows corresponding to actual observation periods in Fig. 3. In that figure a flux change in different bands in the minimum and maximum of the S variability is clearly pronounced. The lightcurve in the U pass band has more steep gradient in comparison to other bands on exit from the minimum of the cycle. On the down grade of the lightcurve in 2013, a new flare (of the F-component) is observed, and it also has more steep gradient in the U pass band.

\begin{figure}[tbh!]
\begin{minipage}[h]{0.18\linewidth}
\center{\includegraphics[width=0.975\linewidth]{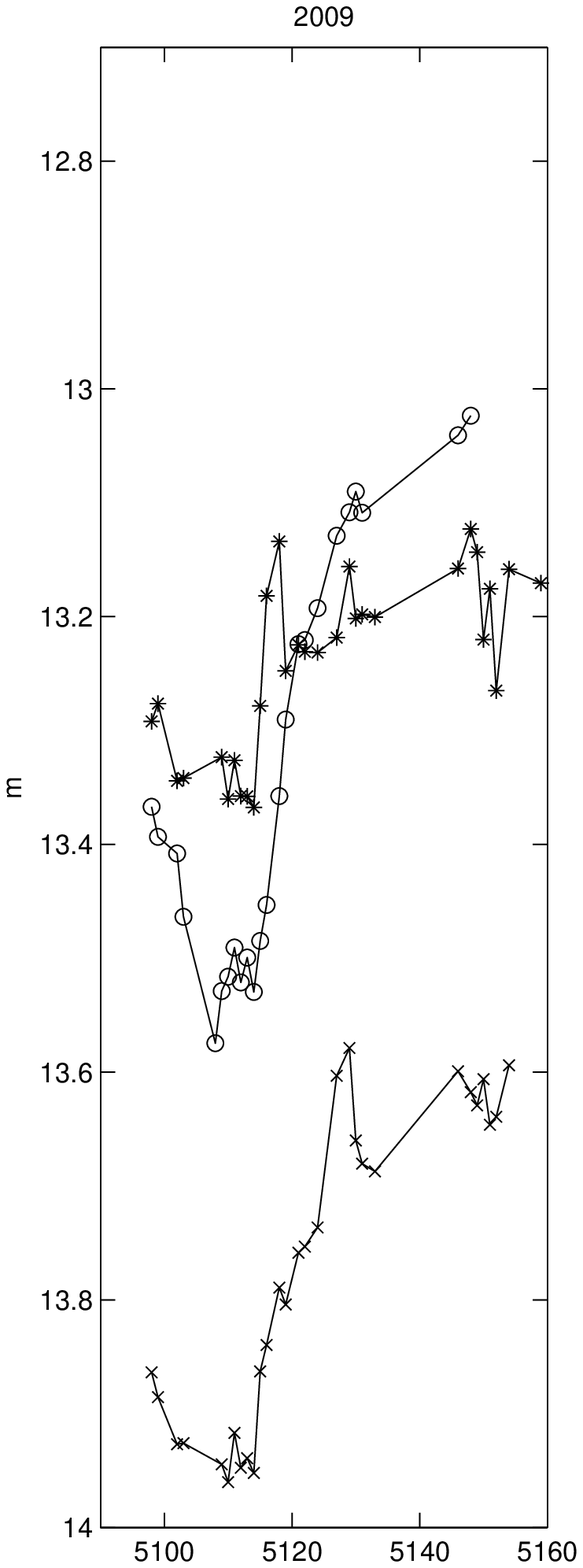}} \\
\end{minipage}
\begin{minipage}[h]{0.18\linewidth}
\center{\includegraphics[width=1\linewidth]{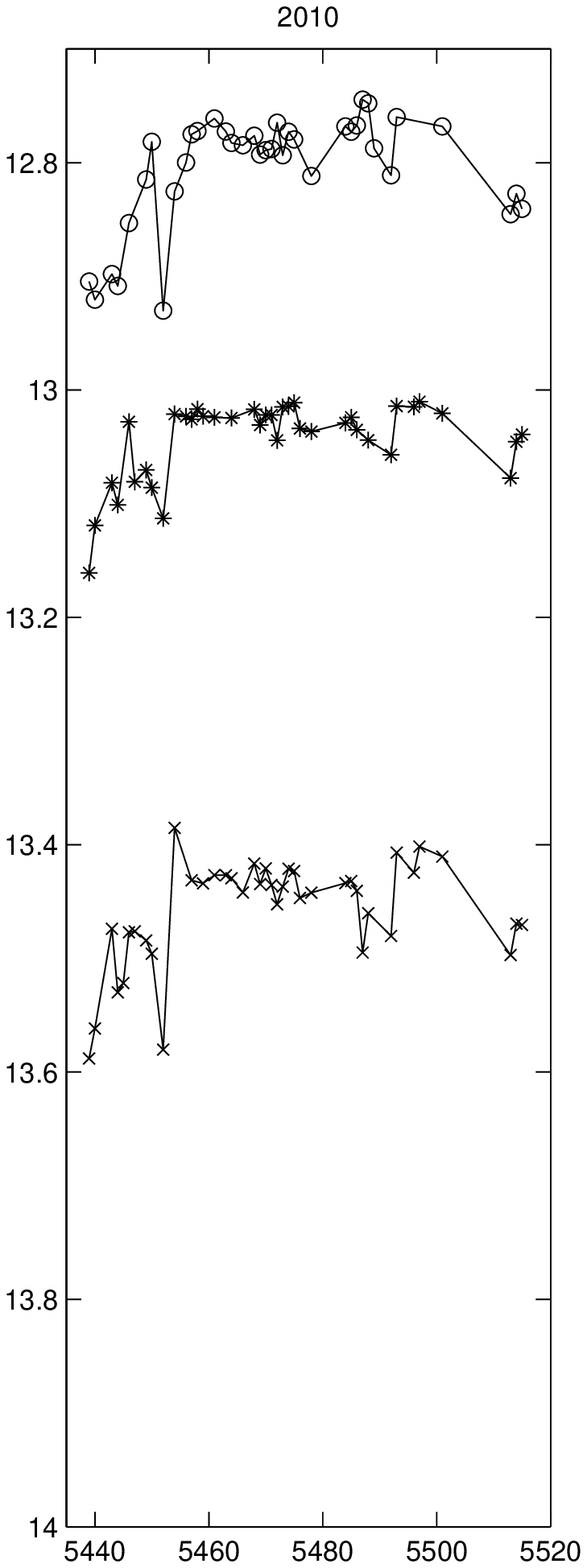}} \\
\end{minipage}
\begin{minipage}[h]{0.18\linewidth}
\center{\includegraphics[width=1\linewidth]{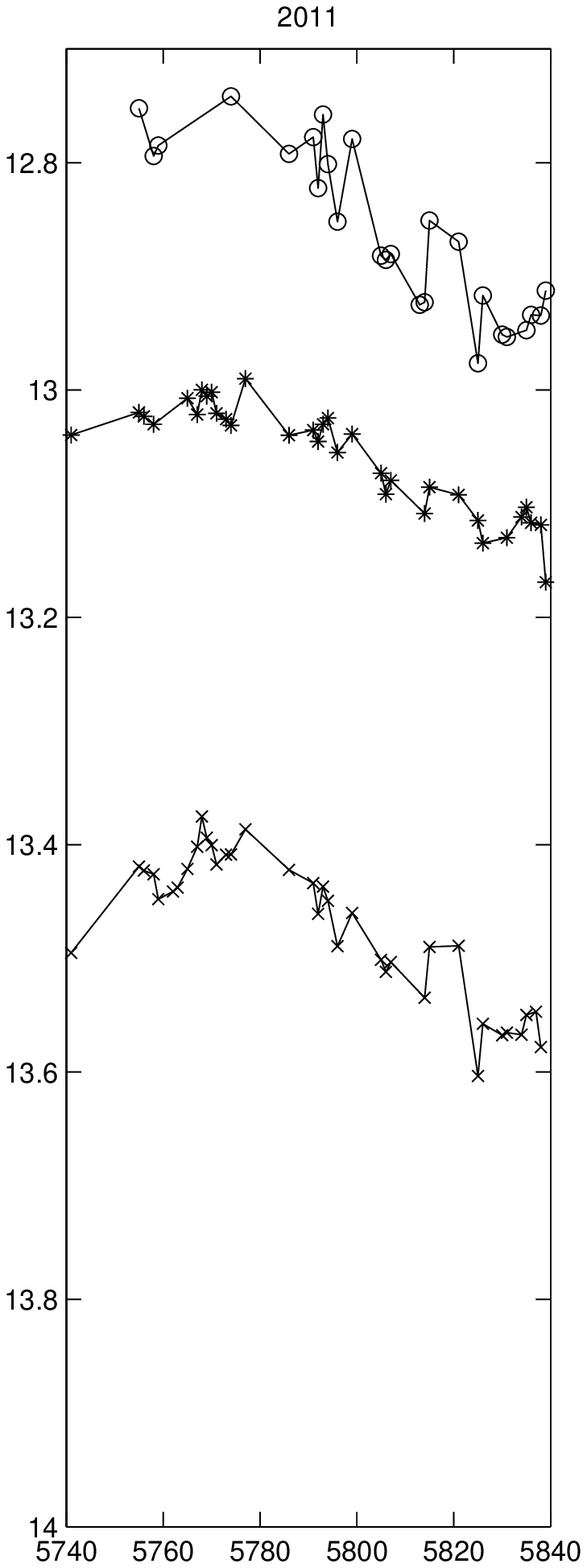}} \\
\end{minipage}
\begin{minipage}[h]{0.18\linewidth}
\center{\includegraphics[width=1\linewidth]{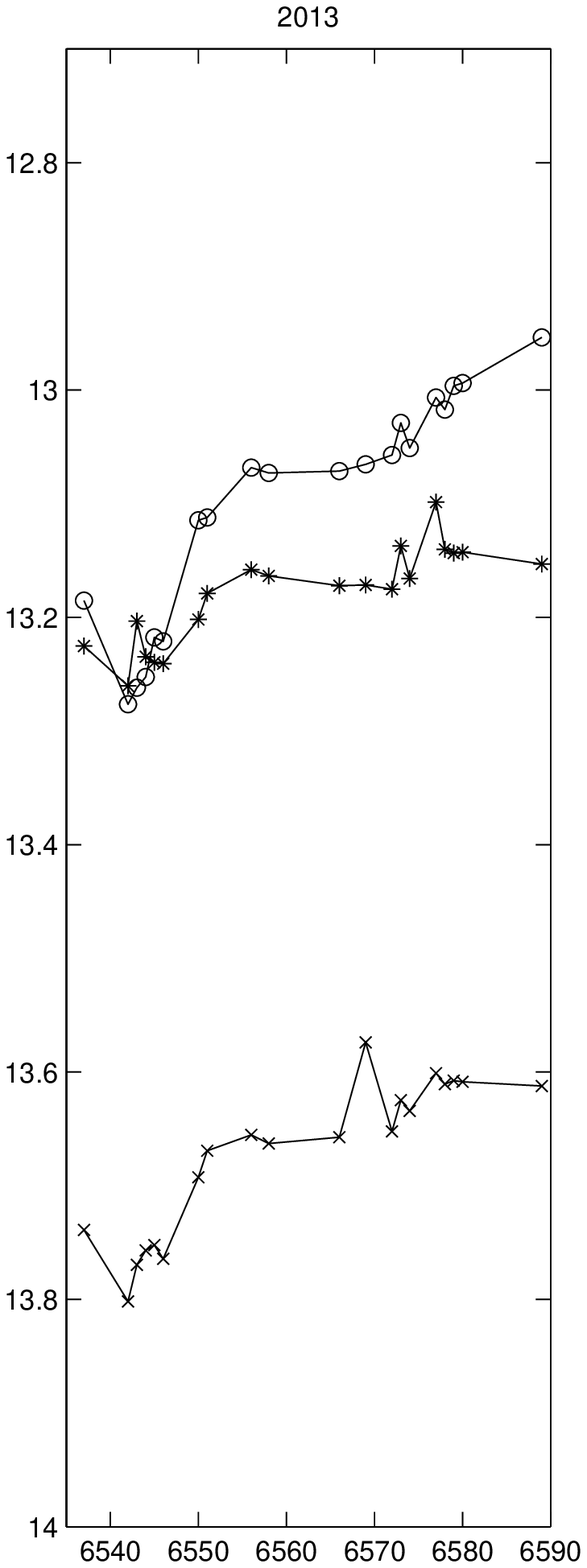}} \\
\end{minipage}
\begin{minipage}[h]{0.18\linewidth}
\center{\includegraphics[width=1\linewidth]{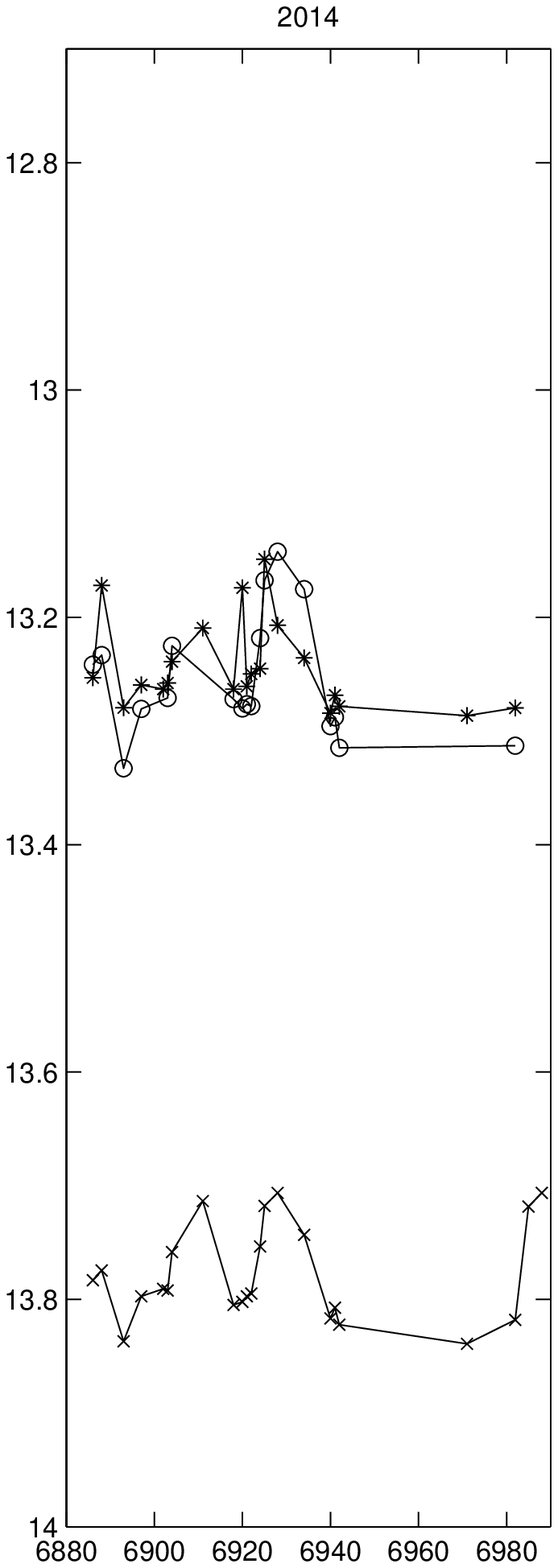}} \\
\end{minipage}
\caption{The light curves of NGC~7469 in actual observation time windows in U (circles), B (crosses), V (stars) in 2009--2014, the aperture $d=10''$. Uncertainties of the photometry is 0.007--0.02 mag, so they could not be presented on the plots.}
\label{20092014}
\end{figure}

Quantitative estimates of the U-B and B-V color indexes behaviour are presented on the two-color diagrams (Fig. 4a,b,c for three apertures $d=10''$, $20''$, $30''$. On the basis of the common assumption that the continuum spectrum of active galactic nuclei is comparable to the black body radiation, we present a color index of the black body radiation vs the black body temperature. The curve is from Straizhis book (1992). The color indexes are for minimum and maximum, ascending and descending branches of the lightcurve, and marked with different symbols (see details in figure legends). Main features of the color index distribution are as follows:

1.	The color becomes bluer along the lightcurves from minimum to maximum of radiation (Fig. 4a-d). Color indexes are situated on the line that is parallel to the black-body radiation curve.

2.	The smaller the aperture is, the bigger the offset of the color indexes to the blue part on the two-color diagram. The same phenomenon for Seyfert galaxies in contrast to normal galaxies was noted e. g. in works by Zasov (1989), Artamonov et al. (2010).

3.  With increase of the aperture, threshold color indexes U-B and B-V change towards lower black-body temperature, and the color becomes redder for the constant part of the galaxy radiation.

The galaxy color change trajectory is shifted with respect to the black-body curve towards the blue part of the diagram. To perform more correct comparison, it is necessary to take into account the galaxy contribution to the radiation of the central part of the galaxy. Furthermore, the radiation can be not quite the black body one. In the framework of the black-body radiation model, the accretion disk temperature in the maximum of the S-component  	approaches almost 8000œK.

\begin{figure}[tbh!]
\begin{minipage}[h]{0.47\linewidth}
\center{\includegraphics[width=1\linewidth]{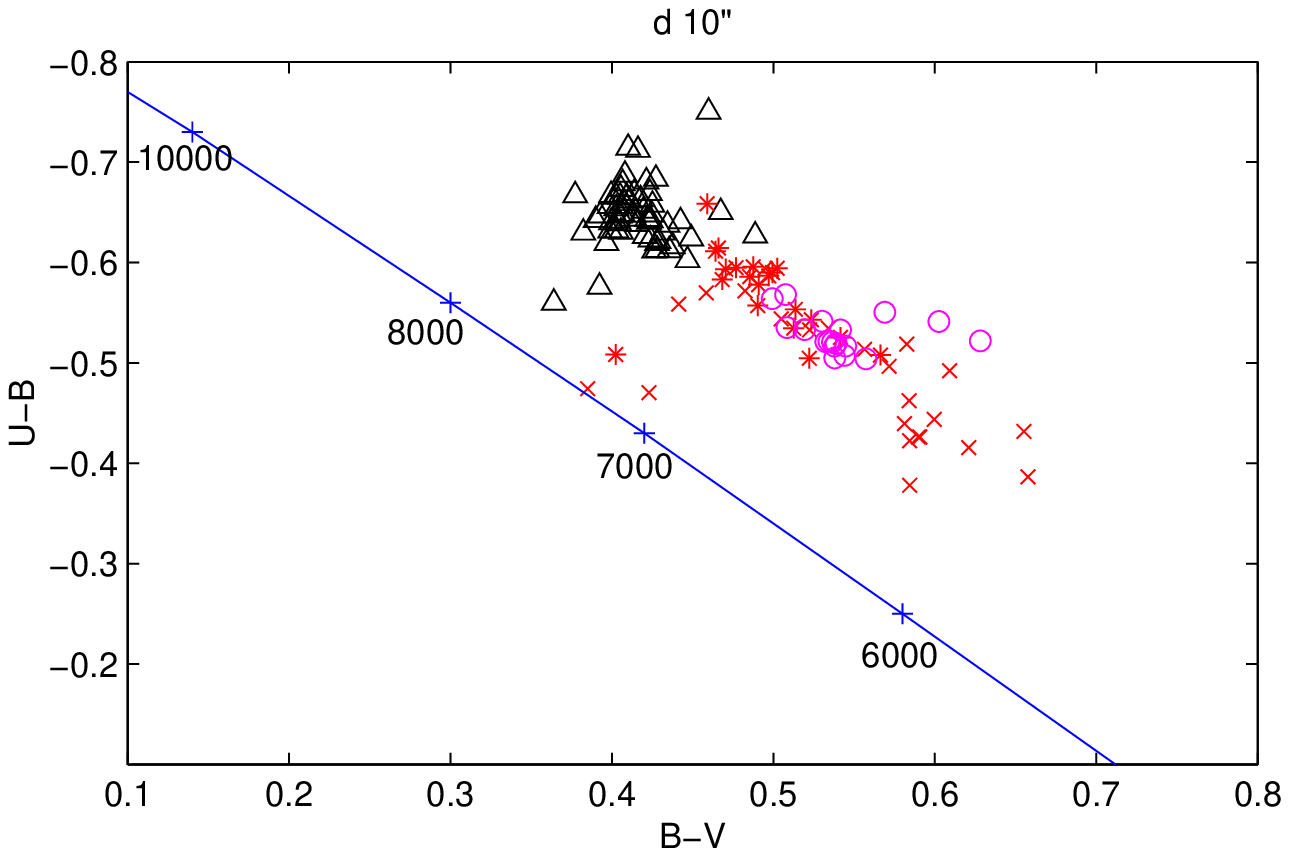}a)} \\
\end{minipage}
\begin{minipage}[h]{0.47\linewidth}
\center{\includegraphics[width=1\linewidth]{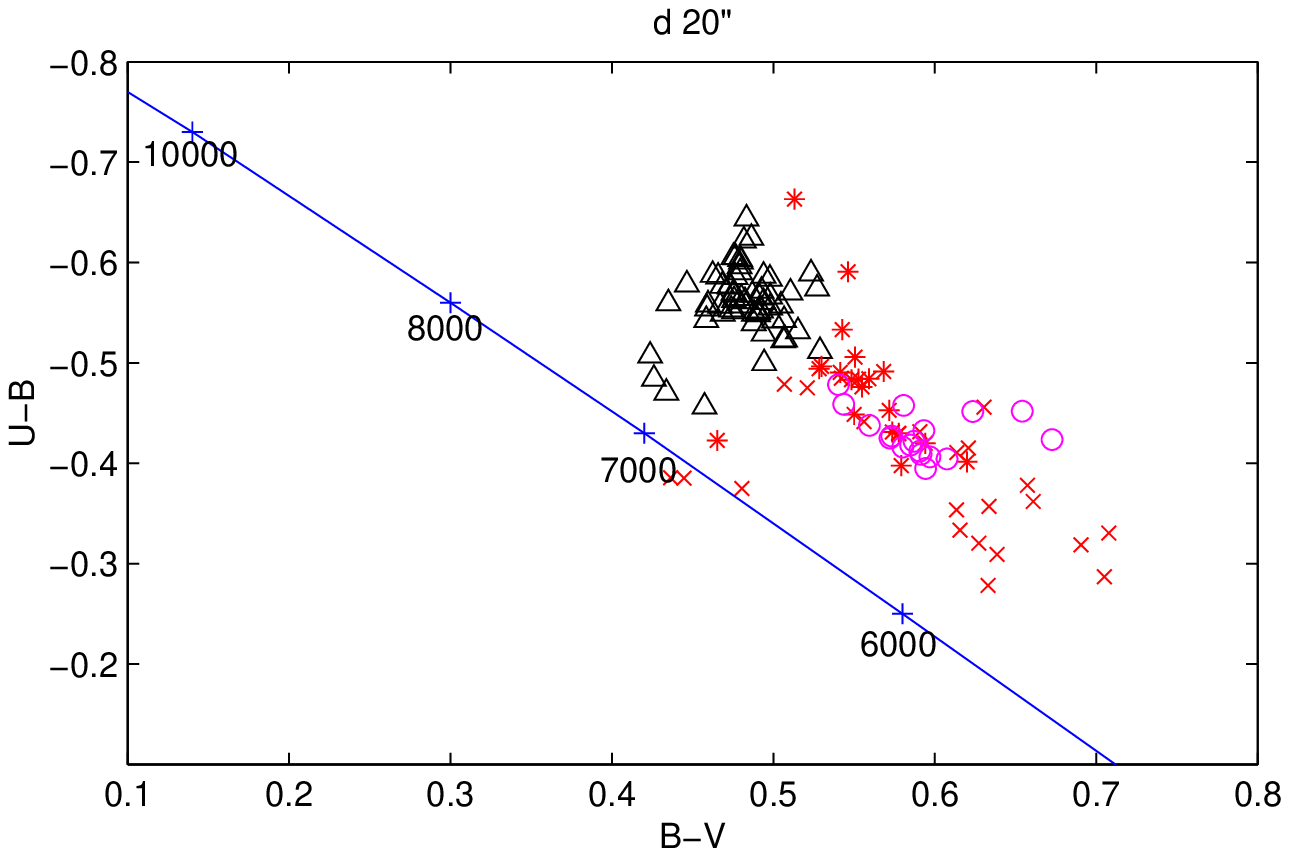}b)} \\
\end{minipage}
\vfill
\begin{minipage}[h]{0.47\linewidth}
\center{\includegraphics[width=1\linewidth]{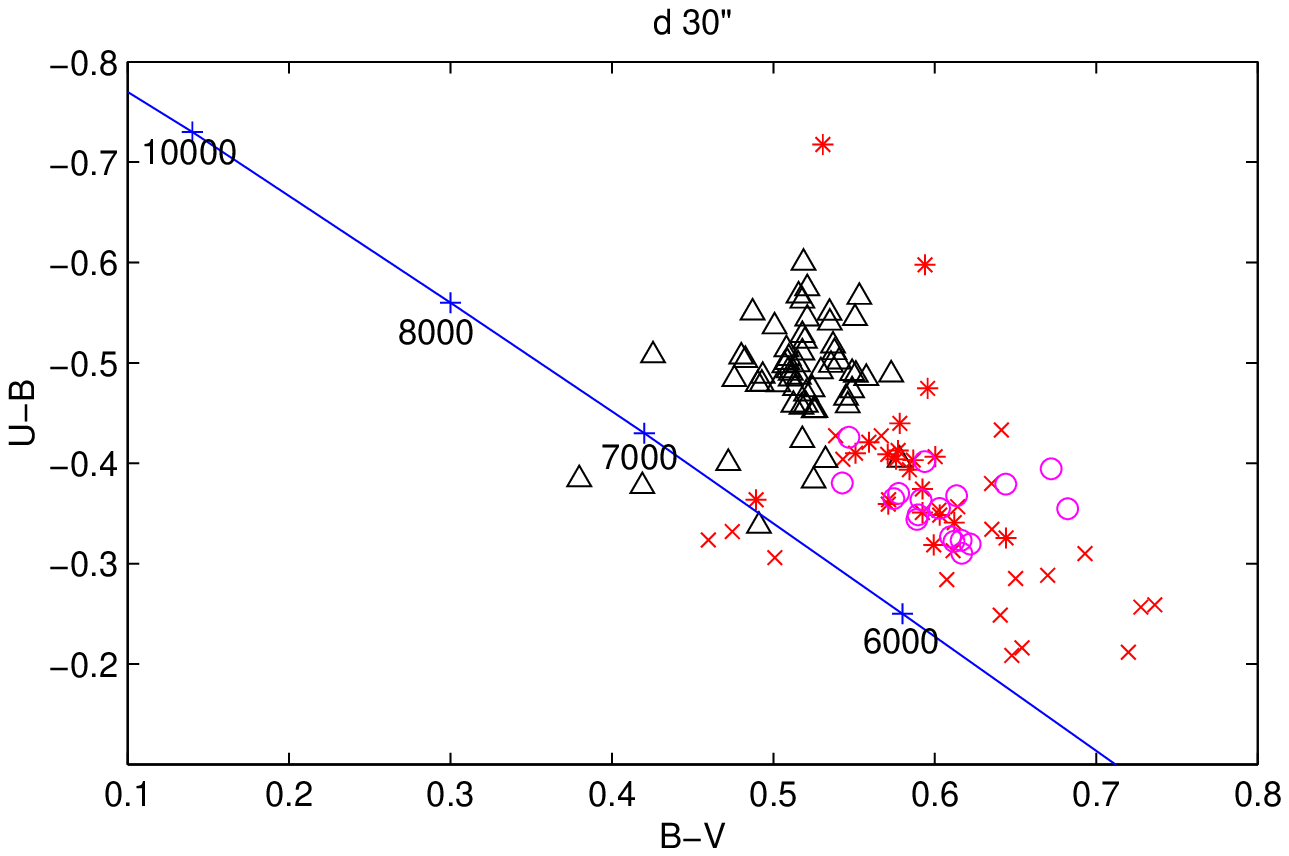}c)} \\
\end{minipage}
\begin{minipage}[h]{0.47\linewidth}
\center{\includegraphics[width=1\linewidth]{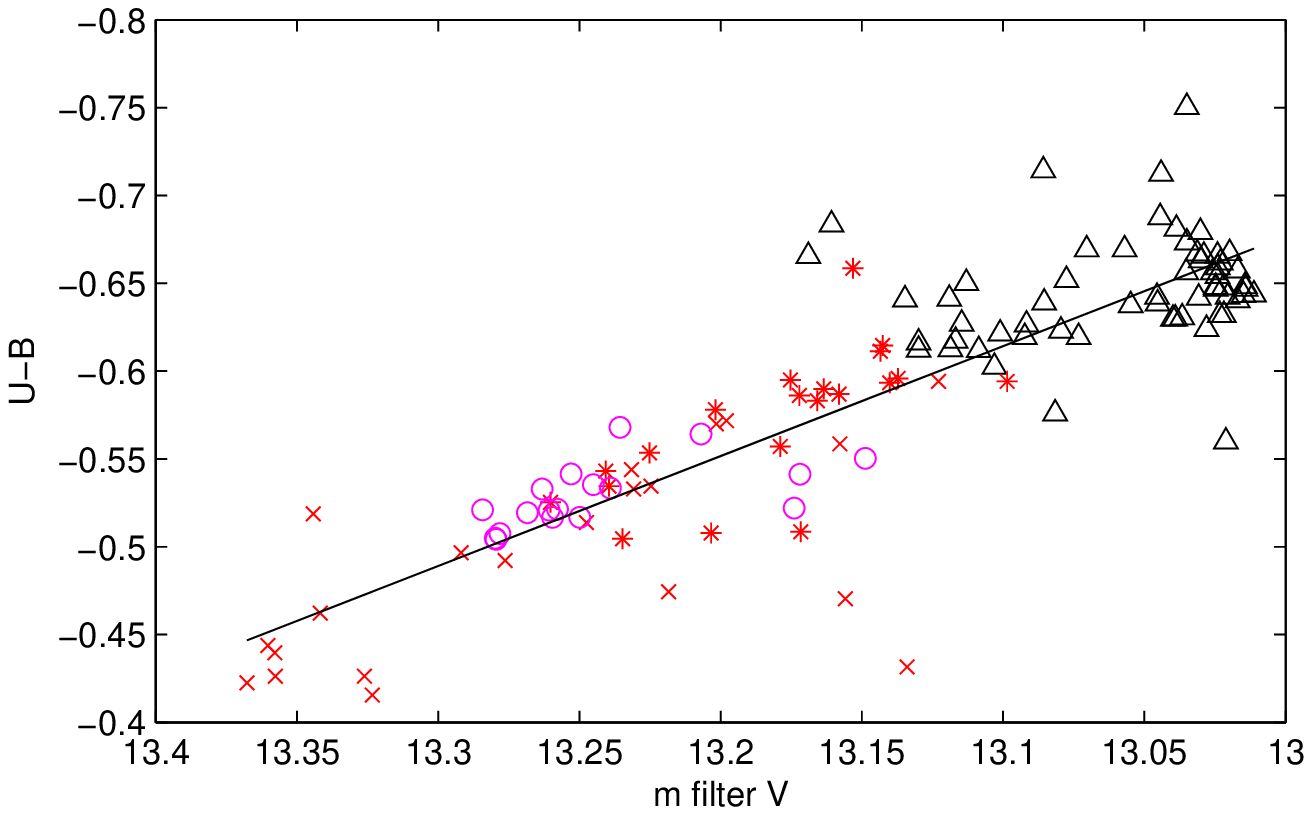}d)} \\
\end{minipage}
\caption{\rm The color diagrams. a)~-- the aperture $d=10''$, b)~-- the aperture $d=20''$, c)~-- the aperture $d=30''$, d)~-- U-B vs V. Crosses~-- 2009 activity minimum, triangles~-- 2010-2011 activity maximum, stars~-- 2013 data, circles~-- 2014 data.}
\end{figure}

\section*{Optical--X-ray correlation}
\noindent
To compare optical and X-ray variability of NGC~7469, we used data from Rossi X-Ray Timing Explorer (RXTE) in the 2-10 keV range (http://cass.ucsd.edu/~rxteagn/NGC7469/NGC7469.html). The RXTE data were obtained for NGC~7469 in 1996 and 2003-2009 (Rivers et al. 2013).

Optical and X-ray lightcurves NGC~7469 in 2009 are presented in Fig.~5, where their similarity can be clearly observed: both have a minimum, an ascending branch and a plateau. The shape of the X-ray lightcurve more
closely coincides with that in U and slightly more poorly with those in B and V, so the X-ray variability better correlates with the optical variability in the U band. In 2009 the optical--X-ray correlation coefficient increases from 0.7 in the I band to 0.9 in the U band. Cross-correlation functions (CCF) were calculated for UBVRI and X-ray data. In this work we calculate the CCFs using the technique which is a modification of the Gaskell-Spark method (1986) and is described in articles by Oknyanskiy (1993), Koptelova et al. (2010), Shimanovskaya et al. (2015). In Fig. 6a, the óóFs for the X-ray (7--10 keV) and optical variability in 2003 based on RXTE data and data from Artamonov et al. (2010) and Doroshenko et al. (2009) are presented. Before CCF calculation we averaged data for every night. In Fig.~6b, CCFs between X-ray data (7-10 keV) and UBVRI data in 2009 are presented.

\begin{figure}[tbh!]
\centerline{
\includegraphics{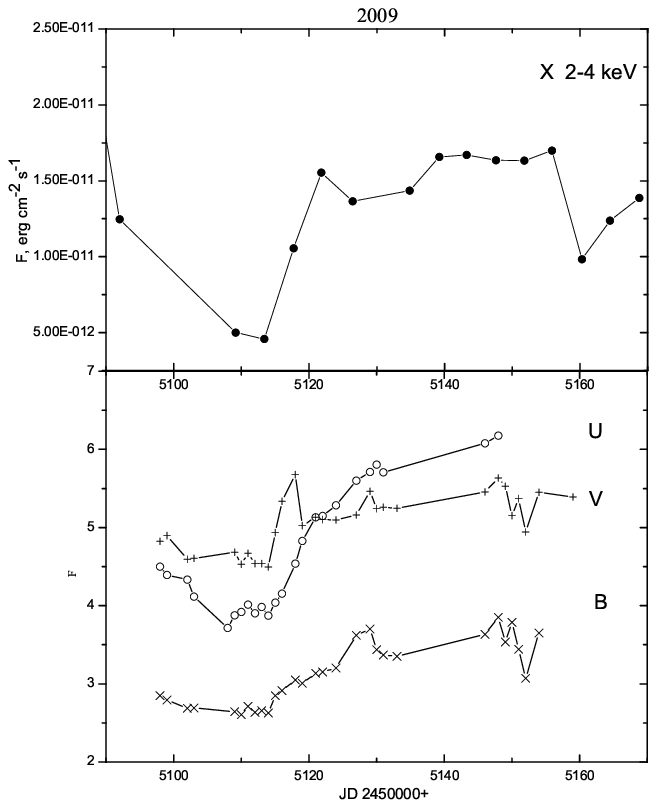}}
\caption{\rm The light curves in 2009. Top~-- X-ray 2-4 keV, bottom~-- U (circles), B (crosses), V (pluses).}
\end{figure}

\begin{figure}[tbh!]
\begin{minipage}[h]{0.47\linewidth}
\center{\includegraphics[width=1\linewidth]{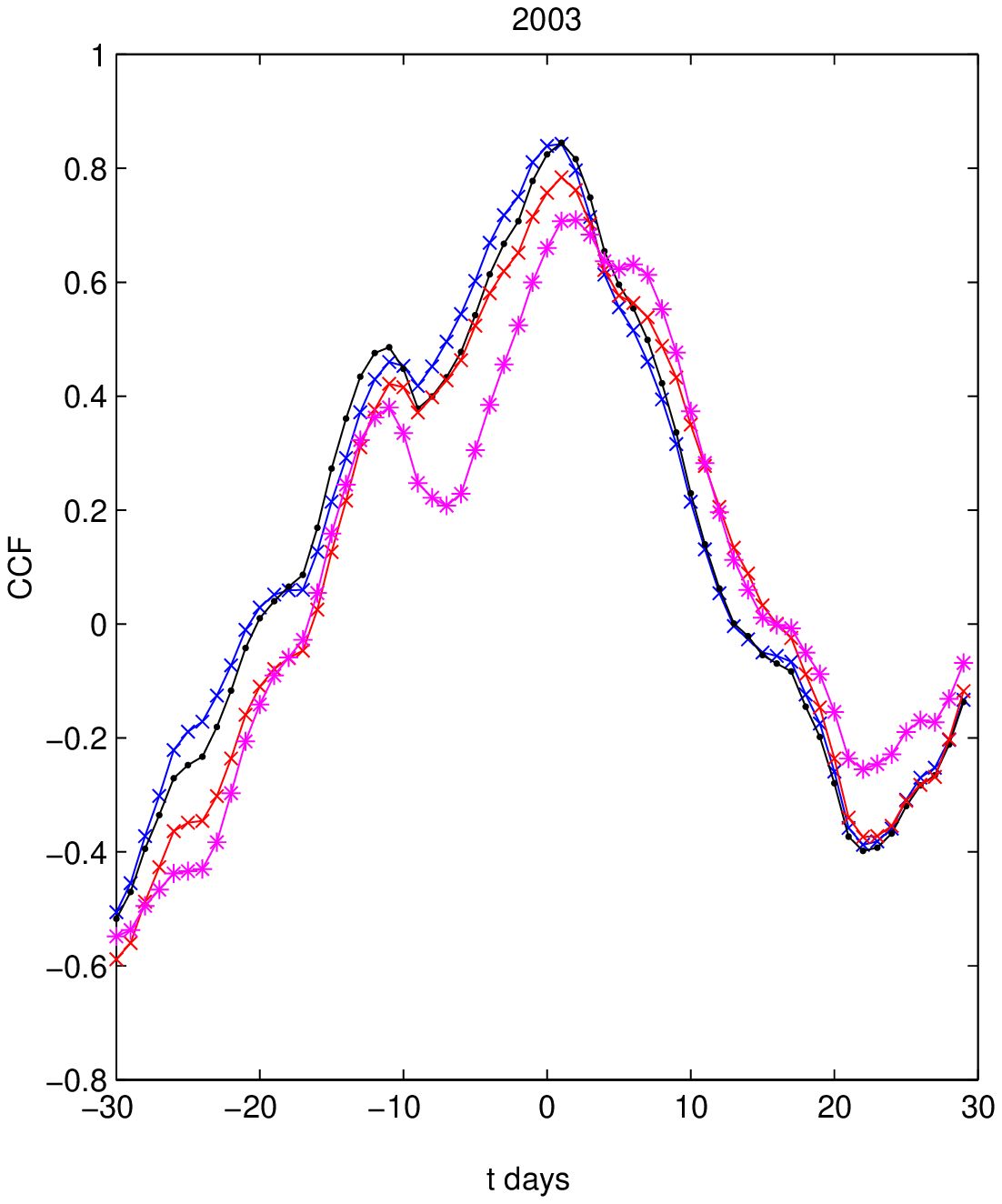}a)} \\ 
\end{minipage}
\begin{minipage}[h]{0.47\linewidth}
\center{\includegraphics[width=1\linewidth]{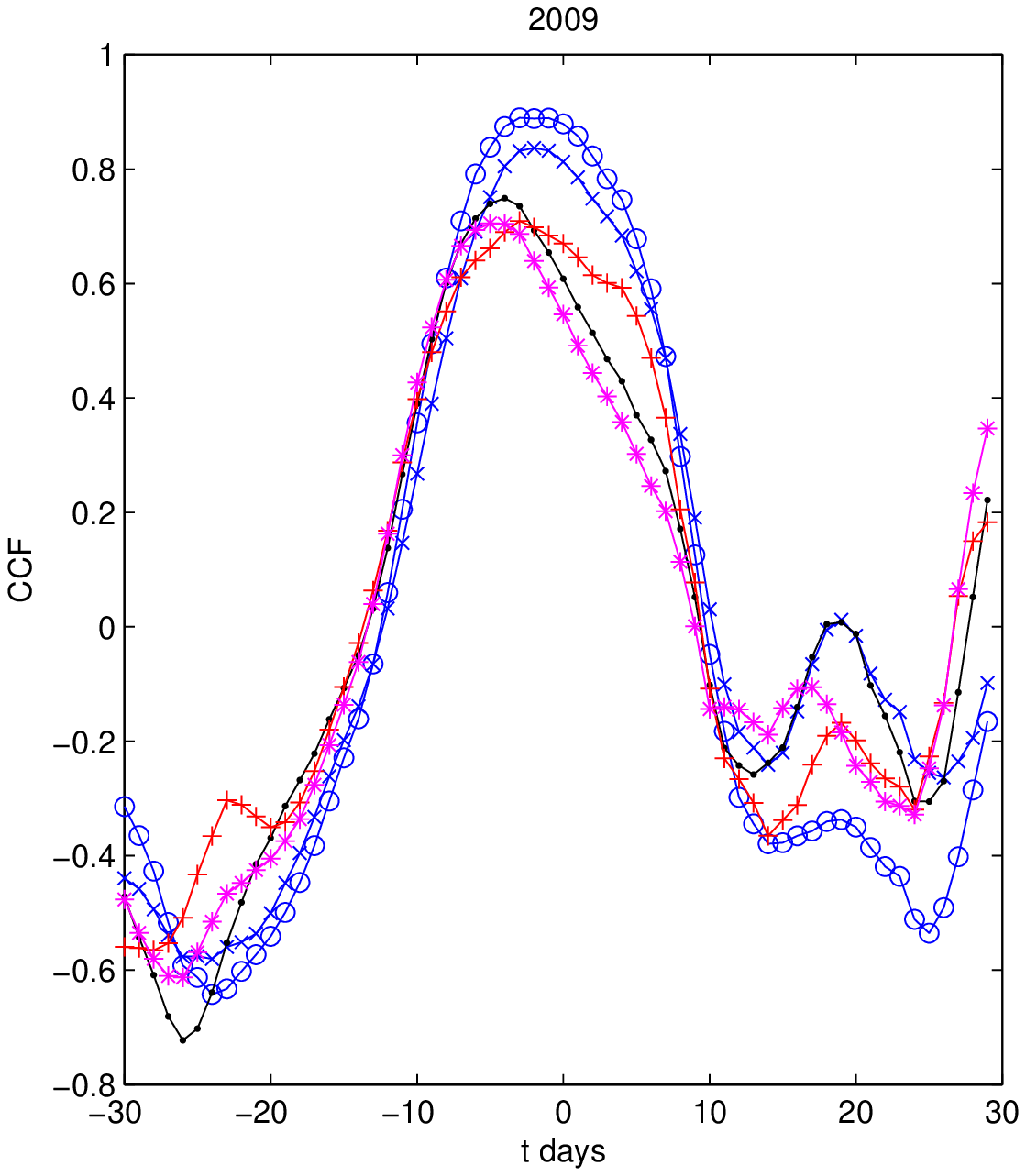}b)} \\
\end{minipage}
\caption{Left: The CCFs of optical and X-ray variability in 2003 (crosses~-- B, dots~-- V, pluses~-- R, stars~-- I). Right: The CCFs of optical and X-ray variability in 2009 (circles~-- U, crosses~-- B, dots~-- V, pluses~-- R, stars~-- I).}
\end{figure}

Comparison of CCFs for 2003 and 2009 reveals that in the NGC~7469 activity minimum (2003) the time delay between X and U is almost zero. The exact value depends on the method that is chosen to calculate it -- centroid or maximum of CCF. But number of correlated pairs is insufficient for even reliable sign determination. 
In 2009, a noticeable  delay of the X-ray variability with respect to the optical one is observed. The delay value increases from the U band to the I band (Fig.~6b), from 2 to 4 days according to preliminary estimates. It is noteworthy that the number of correlated pairs is not sufficient for reliable statistics.


In 2008, a minimum of the NGC~7469 activity is observed. The correlation of the optical radiation (B band) with the X-ray variability is weak ($k\sim 0.6$) for 7-10 keV and even weaker ($k\sim0 .5$) for 2-4 keV range. In 2008, a long-term flare on the lightcurve of NGC~7469 (about two months) is observed, the maximum of that flare occurs in several dozens days after a SN explosion in the galaxy at $~16''$ distance  from its nucleus.

Similar studies were also performed for other active galactic nuclei: Maoz et al.(2000, 2002) investigated connection between optical and X-ray data for NGC 3516. They smoothed lightcurves with a 30 day boxcar running mean and found that the X-ray emission follows the optical one but the deep minimum in the optical variability corresponds to the minimum in the X band with almost zero lag. Observations of NGC 3516 during 5 years (1997--2002) reveals no correlation between optical and X-ray bands except in 1997-1998 when 100 day lag of X-ray behind optical variations was detected. Thus, the correlation is observed when NGC~3516 is in a high state, and  there is no correlation when it is in a low state. Doroshenko et al. (2009) compared variability of the 3ó120 in optical bands (UBVRI) and in the X-ray band (RXTE) for 1996--2008 and found that sometimes the X-ray variability is followed by the optical one, sometimes a zero lag is observed, sometimes the optical variability is leading.

In different activity periods, the optical-X ray correlation is different because of complex and diverse physical nature of processes near the NGC~7469 nucleus. As an addition to well-known reprocessing mechanism, Gaskell (2006, 2007) suggested the anisotropic high-energy emission model for explanation of the optical-X-ray variability and impact of IR dust emission around an accretion disk to a visible band when estimating time delays between optical passbands from B to I. Detailed discussion of processes near the NGC~7469 nucleus can be found in Doroshenko et al. (2010), Chesnok et al. (2009). They compare X-ray luminosity and optical luminosity and conclude that the direct Compton re-processing is not likely to dominate in NGC~7469, it can generate only part of apparent optical luminosity. Other processes, such as the thermal radiation from the accretion disk, the star formation processes, the inverse Compton scattering by hot coronal electrons etc. may also contribute to the optical emission.

\section*{Conclusion}
\noindent

At present, the universally accepted view on the nature of variability in the nuclei of Seyfert galaxies is
an accretion of matter onto a supermassive compact object, and all manifestations of the variability are associated
with the accretion disk. The increase in the duration of the slow component of variability, associated with the accretion disk, indicates possible change in accretion conditions.

In this paper we present results of our UBVRI observations of the Seyfert galaxy NGC 7469 at the Maidanak
Observatory in 2008--2014. We obtained light curves for all passbands with various apertures: $d = 10''$,
$15''$, $20''$, and $30''$.
Analysis of the light curves reveals the presence of a fast variability component with a duration from several days to several tens of days, and a slow component with the duration of about 6 years.
This is the third activity cycle of the NGC 7469 nucleus over the period of its observations at the Maidanak
Observatory from 1990 to 2014.

Our observations showed the following: 

1.  In 2008--2014, the characteristic behaviour of the light curves in all passbands are similar; the differences refer mainly to the variability amplitude.

2. The relative variability amplitude decreases with increasing the wavelength (from U to I) for both components (S
and F). In the observation period, the amplitude of U brightness variations reached its maximum value of 0.9 mag. 

3. The brightening time (ascending branch) is shorter than the fading time (descending
branch) in all passbands. The brightness gradient increases from I to U. The variability amplitude
increases from I to U, indicating a characteristic burst of light in the blue part of the spectrum in active
galactic nuclei. 

4. We analysed the color variations of NGC 7469 in different activity periods. On the (U-B)--(B-V )
diagram we compared the color characteristics of the slow component with the blackbody radiation of a gas
modeling the accretion disk. The color becomes bluer along the light curves from minimum
to maximum light, and the color indices lie on a line that is parallel to the blackbody radiation.
The color of the galaxy NGC~7469 becomes bluer as the nucleus is approached, in contrast to normal galaxies.

5. We analysed the X-ray variability of NGC~7469 in 2008 and 2009 in comparison with the 2003 activity minimum.
In 2008, there is a weak correlation between optical and X-ray variability (the correlation coefficient is 0.5--0.6), which can be explained by an SN Ia explosion that manifests itself in the optical band but does not change the pattern of X-ray variability. In 2009, a U-X correlation with a correlation coefficient that is close to 0.9
is observed. The correlation coefficient and the lag depend on the wavelength in the optical
and X-ray bands. The results of these studies and the results of such an analysis in other papers show
that the correlation between the optical and X-ray variability is different in different activity periods due to the
complex and diverse physics of the processes near the nucleus of a Seyfert galaxy.

\section*{Acknowledgments}

We thank B. Khafizov for the technical support of the Seyfert galaxies monitoring program at the Maydanak observatory. The work is supported by Russian Foundation for Basic Research, the project ID is 14-02-01274.
This work has made use of light curves provided by the University of California, San Diego Center for Astrophysics and Space Sciences, X-ray Group (R.E. Rothschild, A.G. Markowitz, E.S. Rivers, and B.A. McKim), obtained at http://cass.ucsd.edu/$\sim$rxteagn/.


\clearpage

\end{document}